\newcommand{\be}{\begin{equation}}
\newcommand{\ee}{\end{equation}}
\newcommand{\bea}{\begin{eqnarray}}
\newcommand{\eea}{\end{eqnarray}}
\newcommand{\nn}{\nonumber\\}
\newcommand{\p}[1]{(\ref{#1})}

\newcommand{\ve}{\varepsilon}
\documentstyle[12pt]{article}

\begin{document}
\begin{titlepage}
\begin{flushright}
LPENSL-TH-06/2001 \\
hep-th/0111106 \\
November 2001
\end{flushright}
\vskip 1.0truecm
\begin{center}
{\large \bf PARTIAL SUPERSYMMETRY BREAKING AND AdS$_4$ SUPERMEMBRANE}
 \end{center} \vskip
1.0truecm
 \centerline{\bf F. Delduc${}^{\;a,1}$, E.
Ivanov${}^{\;b, 2}$, S. Krivonos${}^{\;b,3}$}
\vskip 1.0truecm
\centerline{${}^{a}$ \it
Laboratoire de Physique ,}
\centerline{\it Groupe de Physique Th\'eorique ENS Lyon,}
\centerline{\it 46, all\'ee d'Italie, F - 69364 - Lyon CEDEX 07}
\vskip5mm
\centerline{${}^{b}$\it Bogoliubov Laboratory of Theoretical
Physics, JINR,}
\centerline{\it Dubna, 141 980 Moscow region, Russia}

\vskip 1.0truecm  \nopagebreak

\begin{abstract}
\noindent
We consider partial spontaneous breaking of $N=1$ AdS$_4$ supersymmetry
$OSp(1|4)$ down to $N=1, d=3$ Poincar\'e supersymmetry in the nonlinear
realizations framework. We construct  the corresponding worldvolume
Goldstone
superfield action and show that it describes the $N=1$ AdS$_4$
supermembrane.
It enjoys $OSp(1|4)$ supersymmetry realized as a field-dependent
modification
of $N=1, d=3$ superconformal symmetry and goes into  the superfield action
of
ordinary $N=1, D=4$ supermembrane in the flat limit. Its bosonic core is
the
Maldacena-type conformally invariant action of the AdS$_4$ membrane. We show
how to reproduce the latter action within a  nonlinear realization of the
AdS$_4$ group $SO(2,3)$. The same universal nonlinear realizations
techniques
can be used to construct conformally-invariant worldvolume actions for
$(d-2)$-branes in generic AdS$_d$ spaces.
\end{abstract}
 \vfill
 \noindent{\it E-Mail:}\\
 {\it 1)francois.delduc@ens-lyon.fr}\\
 {\it 2) eivanov@thsun1.jinr.ru}\\
 {\it 3)krivonos@thsun1.jinr.ru}

 \end{titlepage}

\section{Introduction}
A view of superbranes as theories explicitly exhibiting the
phenomenon of partial spontaneous breaking of global supersymmetry (PBGS)
\cite{BW0,HLP} received a considerable attention
(see, e.g, \cite{BIK,Iv} and references therein).
In the approach with PBGS as the guiding principle, the
manifestly worldvolume supersymmetric  superbrane actions emerge as the
Goldstone superfield actions  associated with nonlinear realizations of some
global space-time supersymmetry groups spontaneously broken down to smaller
supersymmetries. The Goldstone superfields carry the superbrane physical
worldvolume multiplets and are identified with coordinates of some
superspace
of the full supersymmetry. In this sense, superbranes in the PBGS approach
bear
a clear analogy with the  standard nonlinear sigma models which are
nonlinear
realizations of spontaneously broken internal symmetries and describe
Goldstone fields parametrizing the appropriate bosonic coset manifolds.

Until now, the PBGS approach was applied to spontaneously broken  Poincar\'e
supersymmetries in diverse dimensions, in general properly extended by some
central-charge generators. All systems of this kind amount  to $p$- or D$p$-
superbranes on flat Minkowski backgrounds. It is tempting to
generalize the PBGS approach to the case of curved backgrounds.  In most
general setting, this problem amounts to coupling, in a proper way,  the
flat
PBGS actions to the target and worldvolume supergravities, thus implying
the passage from the PBGS concept to that of the partial breaking of local
supersymmetry. As the first step toward  this goal it is natural to
adapt the PBGS approach for constructing worldvolume superfield actions of
superbranes on some special homogeneous  superbackgrounds. In view of the
famous  AdS/CFT correspondence \cite{M,Pol,Witten}, it seems of primary
interest to look, from this point of view, at the superbranes living on the
superextended  AdS$_n$ $\times S^m$-type  backgrounds. The
Green-Schwarz-type
worldvolume  actions for such superbranes were intensively discussed in
literature (see.,  e.g., \cite{GSads0}-\cite{adsd}). To the best of our
knowledge, no relevant worldvolume superfield actions were explicitly
constructed.

In this letter we present the PBGS action for a simple example of AdS
superbranes, the AdS$_4$ supermembrane, and demonstrate that it goes into
the
known PBGS action of the ordinary $D=4$ supermembrane \cite{IK1} in the
limit of
the infinite AdS radius. The bosonic core of the action is a 3-dimensional
analog  of the scale invariant 4-dimensional Maldacena action \cite{M}. We
also
show how this bosonic membrane action can be independently derived from
the appropriate  nonlinear realization of the AdS$_4$ group $SO(2,3)$.
The derivation can be directly extended to the generic case of $(d-2)$-brane
in AdS$_d$.

The AdS (super)groups are realized  as (super)conformal
groups on the (super)spaces the bosonic dimension of which is smaller
by 1 compared to the bulk AdS (super)spaces.
Thus, from the worldvolume perspective,  the corresponding PBGS techniques
should deal with the appropriate nonlinear realizations of superconformal
groups in diverse dimensions. While nonlinear realizations of
superconformal symmetries in superspace were already considered in the
literature \cite{NLRconf}, we shall argue here on the example of the AdS$_4$
supermembrane that relevant to the  PBGS approach is a rather special type
of
them. In the considered case the underlying supersymmetry is the $N=1$
AdS$_4$ one
which eventually takes the form of nonlinearly realized $N=1, d=3$
superconformal symmetry on the worldvolume superspace of the supermembrane.

\section{AdS$_4$ membrane from the coset approach}
To fix our basic assertions, we start with the bosonic case of AdS$_4$
membrane. Whereas it is known how to derive the Nambu-Goto action
for the branes in the $d$-dimensional flat Minkowski background from the
nonlinear realizations (coset) approach applied to relevant Poincar\'e group
\cite{West,Iv1}, no such a self-contained construction was presented so far
for AdS
branes. Here we do this for AdS$_4$ membrane.

The algebra of the AdS$_4$ group $SO(2,3)$ in the $d=3$
spinor $SL(2,R)$ notation and in the basis most appropriate  for our
purposes
reads: \bea && \left[  M_{ab},M_{cd}\right] =\ve_{ac}M_{bd}+\ve_{ad}M_{bc}+
      \ve_{bc}M_{ad}+\ve_{bd}M_{ac} \equiv \left( M\right)_{ab, cd} =  -
\left(
M\right)_{cd, ab}\;, \nn && \left[ K_{ab},K_{cd}\right] = -  \left(
M\right)_{ab,cd} \;,\;  \left[ M_{ab},K_{cd}\right] =
\left( K\right)_{ab,cd}
\;, \;   \left[ M_{ab}, P_{cd}\right] = \left( P \right)_{ab,cd} \;, \nn &&
\left[ K_{ab}, D\right] = -2P_{ab}+2mK_{ab} \;, \;  \left[ P_{ab}, D\right]
=
-2mP_{ab} \;,\;\left[ P_{ab}, P_{cd} \right] = 0~, \nn && \left[
K_{ab},P_{cd}\right] =-2\left( \ve_{ac}\ve_{bd}+    \ve_{bc}\ve_{ad}\right)
D
-m \left( M \right)_{ab,cd} \;. \label{bosads} \eea
Here, $a,b = 1,2$ and the contraction parameter $m$ is proportional to the
inverse AdS$_4$ radius. We use the following conjugation rules
\be
P_{ab}^\dagger =P_{ab}\;, \; M_{ab}^\dagger = -M_{ab}\; ,\;
K_{ab}^\dagger = -K_{ab}\;,\; D^\dagger=D\; , \; m^\dagger = -m \;,
\ee
which allow us to avoid the appearance of the imaginary unit in the
commutation relations. The $SO(1,2)$ generators $M_{ab}$ together with
$K_{ab}$ form the algebra of $D=4$ Lorentz group $SO(1,3)$. In the $m=0$
limit
\p{bosads} goes over into $D=4$ Poincar\'e algebra, with $P_{ab}, D$ forming
the $4$-momentum generator.

Another basis in the algebra \p{bosads}, which makes manifest
its interpretation as $d=3$ conformal algebra, is achieved by passing to
the generators
\be\label{conformalbasis} {\tilde
K}_{ab}=\frac{1}{m}K_{ab}-\frac{1}{2m^2}P_{ab} \quad ,\; {\tilde D} =
\frac{1}{m} D \; ,
\ee
which are the standard $d=3$ special conformal and dilatation generators:
\bea
&& \left[ {\tilde K}_{ab}, {\tilde K}_{cd}\right] =0\; ,\;
 \left[ M_{ab},{\tilde K}_{cd}\right] = \left( {\tilde
K}\right)_{ab,cd}\;,\;
\left[ {\tilde K}_{ab}, {\tilde D}\right] = 2{\tilde K}_{ab}\;, \nn
&& \left[ P_{ab}, {\tilde D}\right] = -2P_{ab}\;,\;
 \left[ {\tilde K}_{ab},P_{cd}\right] =
      -2\left( \ve_{ac}\ve_{bd}+\ve_{bc}\ve_{ad}\right){\tilde D}
  - \left( M \right)_{ab,cd} \;.
\eea
In this conformal basis, any dependence on the dimensionful parameter $m$
disappears.

One more basis can be obtained by passing to the generator
$\hat P_{ab}= P_{ab} - m K_{ab}$. It corresponds to the ``old'' standard
form of the AdS$_4$ algebra (see, e.g., \cite{sorin}), the basic commutation
relation of which can be written as $[P_A, P_B] = m^2M_{AB}~, \;\;A,B= 0,
1,2,3~,$ where $M_{AB} \equiv \left(M_{ab}, K_{cd}\right) $ are generators
of the $D=4$ Lorentz group $SO(1,3)$ and $P_A \equiv \left(\hat P_{ab}, D
\right)$ are generators of the curved $SO(2,3)/SO(1,3)$
translations. In the ``old'' standard basis the AdS$_3$ subalgebra $so(2,2)
\propto \left( \hat P_{ab}, M_{ab} \right)$ of  $so(2,3)$ is manifest. On
the
contrary, in our basis \p{bosads} the $d=3$ Poincar\'e subalgebra $\propto
\left(P_{ab}, M_{ab}\right)$ is manifest (together with the manifest
$so(1,3)$).
basis
The generators $(P_{ab}, D)$ form the maximal solvable subalgebra of
$so(2,3)$. Any AdS$_d$ algebra $so(2, d-1)$ can be written in the basis
where
the $(d-1)$ -dimensional Poincar\'e  symmetry algebra is manifest,
the $(d-1)$-dimensional translation operator together with the dilatation
generator form a solvable subalgebra and also the $d$-dimensional Lorentz
group algebra $so(1,d-1)$ is manifest \cite{solv}.  This basis, the
particular case of which is just \p{bosads}, is indispensable  while
considering AdS branes.

Now we consider the coset $SO(2,3)/SO(1,2)$ in the following
parametrization:
\be\label{bcoset}
g=e^{x^{ab}P_{ab}}e^{q(x)D}e^{\Lambda^{ab}(x)K_{ab}} \;.
\ee
The parameters $x^{ab} = -(x^{ab})^\dagger $ and $q(x) = - q^\dagger(x)$
provide a specific parametrization of the coset $SO(2,3)/SO(1,3) \sim $
AdS$_4$, just adapted to the above solvable-subgroup basis of the
$so(2,3)$-algebra  \p{bosads}. The extra vector parameter
$\Lambda^{ab}(x) = (\Lambda^{ab}(x))^\dagger $ parametrizes the coset
$SO(1,3)/SO(1,2)$. We shall see that its inclusion is imperative for
deducing the AdS$_4$ membrane action from  the coset approach (just like the
case of membrane in a flat background, which corresponds to the nonlinear
realization of the $D=4$ Poincar\'e group in its coset over $SO(1,2)$
\cite{Iv1}
and is the $m=0$ limiting case of our consideration ). Taking into account
that the parameters associated with $P_{ab}$ are the $d=3$  space-time
coordinates, the resulting nonlinear realization actually describes the
spontaneous breaking of $SO(2,3)$ down to its $d=3$ Poincar\'e subgroup. The
latter will finally be the only subgroup realized linearly on
the AdS$_4$ membrane worldvolume $\{x^{ab} \}$.

The full set of the $SO(2,3)$ transformations of the coset parameters in
\p{bcoset} can be found by acting on \p{bcoset} from the left by various
$SO(2,3)$ group elements. For our purposes most interesting are the
$d=3$ conformal transformations of the AdS$_4$ coordinates $(x^{ab}, q(x))$.
They  are generated by the left shift with $g_0=e^{b^{ab}\tilde K_{ab}}$,
where $\tilde K_{ab}$ was defined in \p{conformalbasis}:
\be
\delta x^{ab}=4\left( x^2 b^{ab}-2x^{cd}b_{cd}x^{ab} \right) -
  \frac{1}{2m^2}\,e^{4mq} b^{ab}\;, \quad
\delta q=-{4\over m}x^{ab}b_{ab} \;. \label{confnonl}
\ee
It is important to realize that the algebra of these transformations still
coincides with the $d=3$ conformal group algebra, they simply provide a
different realization of the latter, such that the Goldstone field $q(x)$
proves essentially involved. It is worth mentioning here that the same
generators $P_{ab}, D$ can be regarded to span another coset of $SO(2,3)$,
that over its subgroup $\left(\tilde K_{ab}, M_{ab}\right)$. With this
choice
of the stability subgroup the same coordinates $x^{ab}, q$ carry the
standard
realization  of $SO(2,3)$ as the $d=3$ conformal group: in $\delta x^{ab}$
only the first piece remains, while $q$ still has the same transformation
rule, being a dilaton. Thus different choices of the stability subgroups
give
rise to different realizations. From the perspective of the full
coset $SO(2,3)/SO(1,2)$ \p{bcoset}, these two realizations are related by
nonlinear redefinitions of the coset parameters, such that $x^{ab}, q(x)$
and
$\Lambda^{ab}(x)$ are mixed up in a non-trivial way. It turns out that just
the
separation of the coset parameters into $d=3$ space-time coordinates and
Goldstone fields as in  \p{bcoset} immediately leads to the correct AdS$_4$
action. Correspondingly, it is just the field-dependent $d=3$ conformal
transformations \p{confnonl} under which this action proves to be invariant.

The basic ingredients in constructing the action are left-invariant Cartan
one-forms. In the obvious notation, they are defined by the standard
relation
\be g^{-1}dg = \omega_P\cdot P + \omega_D D + \omega_K\cdot K +
\omega_M\cdot
M ~.
\ee
As usual, the forms associated with the coset generators are transformed
homogeneously, while the $so(1,2)$ Cartan form $\omega_M^{ab}$
has an inhomogeneous transformation rule. For our purposes it is enough to
know
the explicit structure of the following two coset space Cartan forms
\bea\label{bcartanf} && \omega^{ab}_P = e^{-2mq}\left( dx^{ab} +
\frac{4\lambda^{ab}\lambda_{cd}dx^{cd}}{1-2\lambda^2}\right)
+\frac{2\lambda^{ab}dq}{1-2\lambda^2} \equiv E^{ab}_{cd}dx^{cd}\;,
\label{Pform} \\
&& \omega_D= \frac{1+2\lambda^2}{1-2\lambda^2}\left(  dq+
\frac{4e^{-2mq}\lambda_{ab}dx^{ab}}{1+2\lambda^2}\right) \;,\label{Dform}
\eea
where
\be
\lambda^{ab}\equiv
\frac{\tanh\sqrt{2\Lambda^2}}{\sqrt{2\Lambda^2}}\Lambda^{ab}\;, \; \lambda^2
=
\lambda^{ab}\lambda_{ab} \;.  \ee

{}From the structure of \p{Dform} it is clear that $\lambda^{ab}$ can be
covariantly expressed through $q(x)$ by the inverse Higgs \cite{invh}
constraint
\be
\omega_D = 0 \quad \Rightarrow \quad
\frac{4\lambda_{ab}}{1+2\lambda^2}= -e^{2mq}\partial_{ab}q
\quad
\Rightarrow \quad \lambda_{ab} = -{1\over 2} e^{2mq}
\frac{\partial_{ab}q}{1 +\sqrt{1 -{1\over 2}e^{4mq}(\partial
q)^2}}\;.\label{lambdaeq} \ee
Then the dreibein in \p{Pform} takes the simple form
\be
E^{ab}_{cd} = e^{-2mq}\delta^{(a}_{(c}\delta^{b)}_{d)} - {1\over
2}e^{2mq}\frac{1} {1 +\sqrt{1 -{1\over 2}e^{4mq}(\partial
q)^2}}\partial^{ab}q  \partial_{cd}q~. \label{3bein}
\ee
The simplest appropriate invariant is the covariant volume of the $d=3$
space,
$$
\int d^3x\, \mbox{det}\,E(q)~,
$$
and the correct invariant action vanishing when $q = 0$ is given by (up to a
normalization factor)
\be
S = \int d^3x \;\left[e^{-6m q} - \mbox{det}\, E(q)\right] =
 \int d^3x\; e^{-6mq} \left(1 -\sqrt{1-\frac{1}{2}e^{4mq}\partial^{ab}q
  \partial_{ab}q}\,\right) \;. \label{baction2}
\ee
By construction, it possesses all symmetries of the AdS$_4$ space and in
the limit $m=0$ goes into the standard static-gauge form of the Nambu-Goto
action for membrane in four-dimensional Minkowski space. Note that the term
$\sim\int d^3x e^{-6mq} $ is invariant  under the nonlinear
$d=3$ conformal transformations \p{confnonl} and dilatations on its own
right.

Another, equivalent way to deduce the same action \p{baction2} is to start
from the dreibein in \p{Pform} with $q$ and $\lambda_{ab}$ as independent
fields. Then the invariant action takes the form
\be\label{baction1} S = \int d^3x\; \left[e^{-6mq} - \mbox{det}\, E(q,
\lambda)\right] = -2\int d^3x\; e^{-6mq}\left( \frac{2\lambda^2+
e^{2mq}\lambda^{ab}\partial_{ab}q}{1-2\lambda^2}\right) \;. \ee
Equation of motion for $\lambda^{ab}$ yields just the inverse Higgs
expression \p{lambdaeq}, and substituting it back into \p{baction1}
brings the latter into the form \p{baction2}.

To see that the action \p{baction2} indeed describes a membrane embedded
into the AdS$_4$ background, let us look at the induced distance defined as
the square of $\omega_P^{ab}$ with the dreibein \p{3bein}
\be
ds^2 = \omega^{ab}_P\omega_{P\,ab} = e^{-4mq}\, (dx^{ab}dx_{ab})- {1\over 2}
dq
dq~. \label{dist}
\ee
Introducing $U = e^{-2mq}$ and rescaling $x^{ab} = {1\over 2\sqrt{2}m}\tilde
x^{ab}$, one  can rewrite  \p{dist} and \p{baction2}, up to some overall
constant factors, in the form
\bea
ds^2 = U^2\, (d\tilde x^{ab}d\tilde x_{ab}) - \left(\frac{dU}{U}\right)^2~,
\quad   S = \int d^3\tilde x \;U^3\; \left(1 -\sqrt{1 -
\frac{(\tilde\partial
U\cdot \tilde\partial U)}{U^4}}\,\right)~.
\label{standr} \eea
Thus $ds^2$ is
recognized as the $d=3$ pullback of the standard invariant interval on
AdS$_4$
in the parametrization which is of common use in the  AdS/CFT literature,
while $S$ as the $d=3$ analog of the Maldacena scale-invariant brane action
on
AdS$_5$ \cite{M} (actually, of the scalar fields  piece of his full D3-brane
action). The derivation of this form of  the AdS$_4$  interval from the
coset
$SO(2,3)/SO(1,3)$ parametrized by coordinates associated with the solvable
subgroup generators (and a generalization to the generic case of AdS$_d$),
as
well as deducing the field-dependent conformal transformations
\p{confnonl}, were given  in \cite{solv} (see also \cite{padua}). To
our knowledge, the explicit derivation of the AdS$_4$ membrane action from
the coset approach we have given here is new. It can be straightforwardly
extended to the case of $(d-2)$-brane in AdS$_d$ in a static gauge for
an arbitrary dimension $d$ \cite{GSads0,adsd}. In the generic case, the
only basic Goldstone field is also dilaton $q(x)$, while an analog of the
extra 3-dimensional coset factor $SO(1,3)/SO(1,2)$ in \p{bcoset} is the
$(d-1)$-dimensional coset $SO(1,d-1)/SO(1,d-2)$, with the parameters
basically becoming $x$-derivatives of the dilaton after employing the
inverse Higgs constraints. The conformally-invariant $(d-2)$-brane action
is an obvious modification of \p{baction2}, \p{standr} adapted to
the $(d-1)$-dimensional worldvolume.

The mixed $\lambda~, q$ representation for the membrane action \p{baction1}
implies a new interesting (and rather strange) type of duality seemingly
specific just for the AdS (super)branes. To reproduce the standard AdS$_4$
membrane action \p{baction2},  we eliminated $\lambda_{ab}$ by its algebraic
equation of motion. On the other hand, $q$ is also an unconstrained field
and
we can firstly vary \p{baction1} with respect to it, with the result
\be
\partial_{ab}F^{ab} = 6 m\,e^{-2mq}\,\left(1 -\sqrt{1+ 2F^2}\right)~, \quad
F^{ab} \equiv \frac{2\,\lambda^{ab}}{1 - 2\lambda^2}~. \label{qviaF} \ee
In the flat limit $m=0$ this equation becomes $\partial_{ab}F^{ab} = 0$ and
can be interpreted as the Bianchi identity for a $d=3$ Maxwell strength
$F^{ab}$. After  substitution of $F^{ab}$ expressed through the $d=3$ gauge
potential back into the $m=0$ form of \p{baction1}, the latter becomes just
the
$d=3$ Born-Infeld  action \cite{Iv1}, thus displaying the well-known $d=3$
duality between the  membrane Nambu-Goto action in a static gauge and the
$d=3$
Born-Infeld action \cite{Tseyt}. The situation radically changes in the
$m\neq
0$ case: \p{qviaF}  does not longer impose any differential constraint on
$F^{ab}$ and should be rather regarded as the equation expressing $q$
through
$F^{ab}$:  \be
e^{-2mq} = {1\over 6m}\,\frac{(\partial\cdot F)}{1 -\sqrt{1 + 2 F^2}}~.
\ee
In terms of the independent $d=3$ vector field $F^{ab}(x)$, the action
\p{baction1} takes the form
\be
S = -{1\over 2}\left({1\over 6m} \right)^3\int d^3 x\,
\frac{\left(\partial\cdot F\right)^3}{\left(1 -\sqrt{1 + 2 F^2}\right)^2}~.
\label{stract}  \ee
Thus, instead of the familiar NG - BI duality of the flat case, in the
AdS$_4$
case the membrane with the scale-invariant action \p{baction2} proves to be
dual (in the above sense) to some non-gauge $d=3$ vector field theory with
the
strange action \p{stract} which is singular in the limit $m\rightarrow 0$.
For
the time being, the meaning of such a theory is unclear for us. We hope to
say
more on it elsewhere.

\setcounter{equation}0
\section{AdS$_4$ supermembrane}
Our starting point will be the $N=1$ AdS$_4$ superalgebra $osp(1|4)$, once
again in the basis which is a natural generalization of \p{bosads} and is
most convenient for our purposes
\bea\label{n1ads}
&& \left\{ Q_a, Q_b\right\} =2P_{ab}\; ,\;
 \left\{ S_a, S_b\right\} =2P_{ab}-4m\,K_{ab}\; ,\;
 \left\{ Q_a, S_b\right\} =2\ve_{ab}D-2mM_{ab}\; ,\nn
&& \left[  M_{ab},Q_{c}\right] =\ve_{ac}Q_{b}+\ve_{bc}Q_{a} \equiv
  \left( Q\right)_{ab,c} \;, \;
\left[  M_{ab},S_{c}\right] =\left( S\right)_{ab, c}\;, \nn
&&\left[  K_{ab},Q_{c}\right] =\left( S\right)_{ab,c} \;,\;
 \left[  K_{ab},S_{c}\right] =-\left( Q\right)_{ab,c} \;,\nn
&& \left[ P_{ab}, Q_c\right] =0 \;, \;
\left[ P_{ab}, S_c\right] =-2m\left( Q\right)_{ab,c}\;, \;
\left[ D, Q_a\right]=m\,Q_a\;,\;\left[ D, S_a\right]= -m\,S_a\;.
\eea
The bosonic generators are the same as in the previous Section. The
generators
$Q_a, P_{ab}, M_{ab}$ form $N=1, d=3$ super Poincar\'e algebra. The
generators $S_a$ in this basis have the same dimension as $Q_a$ ($Q_a =
(Q_a)^\dagger $,  $S_a = (S_a)^\dagger $). The passing to the conformal
basis, besides the redefinitions \p{conformalbasis}, implies the rescaling
$S_a = m\tilde S_a$, such that the generator $\tilde S_a$ has the
dimension opposite to  $Q_a$ and so has the natural meaning of the $d=3$
conformal supersymmetry generator. The advantage
of the basis \p{n1ads} is that  it manifests the $N=1, d=3$
super Poincar\'e subalgebra of $osp(1|4)$ and still yields the  $N=1, D=4$
super Poincar\'e algebra (in the $d=3$ notation)  in the contraction limit
$m=0$. The $N=1, d=3$ Poincar\'e supertranslations subalgebra
$\propto (Q_a, P_{ab})$ together with the generator $D$ form the maximal
solvable subalgebra of $osp(1|4)$.

We wish to construct a $OSp(1|4)$ extension of the AdS$_4$ membrane
action \p{baction2}, such that it possesses a manifest $N=1, d=3$
supersymmetry  extending the manifest $d=3$ Poincar\'e worldvolume
invariance
of \p{baction2}, and becomes the known action of the flat $N=1, D=4$
supermembrane \cite{IK1} in the contraction limit $m=0$. Just like the
latter
action  is the Goldstone superfield action for the $1/2$ breaking of the
$N=1,
D=4$  Poincar\'e supersymmetry down to the $N=1, d=3$ one, the action we
are seeking for is expected to be a Goldstone superfield action describing
the $1/2$ spontaneous breaking of the $OSp(1|4)$ supersymmetry down to its
$N=1, d=3$ super Poincar\'e subgroup.

The construction of the AdS$_4$ superbrane action is not so straightforward
as in the bosonic case. Already in the case of flat $N=1, D=4$ supermembrane
\cite{IK1} the correct action is by no means a covariant supervolume
of the $N=1, d=3$ superspace. The corresponding superfield Lagrangian
density
is not a tensor, but it is rather of Chern-Simons or WZW type, since it is
shifted by a full derivative under the nonlinearly realized half of full
supersymmetry.  At present, the only known way of constructing such
Goldstone
superfield actions is to start from a {\it linear} realization of the
partially broken supersymmetry in some appropriate superspace. The nonlinear
realization is then recovered by  imposing proper covariant constraints on
the
corresponding superfields  (see, e.g., \cite{BG,RT}). The correct
Goldstone superfield actions arise from some simple superfield invariants of
the initial linear realization after enforcing these  constraints.
There is a systematic  way of searching for such covariant
constraints \cite{DIK,IKZL}. It is a generalization of the analogous
approach worked
out for the case of the totally broken supersymmetry in \cite{IKap}. Now we
are
going to show that these techniques, applied earlier  in PBGS systems with
rigid Poincar\'e supersymmetries, work fairly  well  also in the curved case
at hand and give rise to the sought PBGS action of the AdS$_4$
supermembrane.

As a first step we need to define the appropriate analog of the
aforementioned
linear realization. It turns out that in the AdS case it is already
a sort of nonlinear realization, but with weaker nonlinearities compared to
the
final nonlinear realization which underlies the AdS$_4$ supermembrane
action.

As a natural superextension of the bosonic coset element \p{bcoset} we
choose
the following one:
\be\label{scoset}
g=e^{x^{ab}P_{ab}}e^{\theta^aQ_a}e^{\psi^a
S_a}e^{u(z)D}e^{\Lambda^{ab}(z)K_{ab}} \;.
\ee
Here, the parameters $z \equiv \left(x^{ab}, \theta^a, \psi^a \right)$ are
$N=2, d=3$ superspace coordinates, while $u = u(z)$ and $\Lambda^{ab}(z)$
are
Goldstone superfields given on this superspace. The subspace spanned by the
coordinate  set $\zeta \equiv \left(x^{ab}, \theta^a \right)$ is the
standard
flat $N=1, d=3$ superspace in which $N=1, d=3$ Poincar\'e
supertranslations $\propto (Q_a, P_{ab})$ are realized in a standard way:
\be\label{mansusy}
\delta x^{ab}=a^{ab}-\frac{1}{2}\left( \epsilon^a\theta^b+
   \epsilon^b\theta^a \right)
\; ,\quad \delta\theta^a=\epsilon^a \;.
\ee
These transformation laws are obtained by acting on \p{scoset} from the left
by the group element $g_0=e^{a^{ab}P_{ab}}e^{\epsilon^aQ_a}$.

The rest of the $OSp(1|4)$ transformations except for the $SO(1,2)$
rotations
is nonlinearly realized on the coset coordinates, mixing the $N=2$
superspace coordinates with the Goldstone superfield $u(z)$. Acting on
\p{scoset} from the left by different supergroup elements, it is easy to
find
the explicit form of these transformations.\\

\noindent Broken supersymmetry: $g_0=e^{\eta^aS_a}$,
\bea\label{brokensusy}
&&\delta x^{ab} = 2m\left( \theta^a x^{bc} +\theta^b x^{ac}\right)\eta_c
+\frac{1}{2}e^{4m u}\left( \psi^a\eta^b+\psi^b\eta^a\right)+
\frac{3}{2}me^{4mu}\psi^2\left(\theta^a\eta^b+\theta^b\eta^a\right), \nn
&& \delta\theta^a=4mx^{ac}\eta_c+m\theta^2 \eta^a -3me^{4m u}\psi^2\eta^a\;,
\quad \delta u =2 \theta^a\eta_a \;, \nn
&&\delta\psi^a=\eta^a-2m\left( \eta^b\theta_b \psi^a-
 \eta^a\theta^b \psi_b -  \eta^b\theta^a \psi_b \right).
\eea
Dilatations: $g_0=e^{\alpha D}$,
\be
\delta x^{ab}=2\alpha m x^{ab}\; ,\;
\delta \theta^a=\alpha m \theta^a \; ,\; \delta\psi^a=-\alpha m \psi^a\;,\;
\delta u =\alpha \;.
\ee
As follows from \p{n1ads}, all bosonic transformations (including the
dilatations) are actually contained in the closure of the supersymmetry
transformations, so it is not necessary to explicitly quote them here.

Covariant derivatives of the Goldstone superfield $u(z)$ can be constructed
by the supercoset element \p{scoset} following the generic
guidelines of the nonlinear realizations method. Of actual need for us
will be spinor covariant derivatives. Without entering into details, these
are as follows
\bea
&&{\tilde\nabla}_a^Q u =\frac{1}{\sqrt{1-2\lambda^2}}\left( {\nabla}_a^Q-
  2\lambda_a^b{\nabla}_b^S\right)u\;, \;
{\tilde\nabla}_a^S u=\frac{1}{\sqrt{1-2\lambda^2}}\left( {\nabla}_a^S+
  2\lambda_a^b{\nabla}_b^Q\right)u\;, \label{fullcovder} \\
&& \nabla^Q_a u = e^{mu}\left( D_a -
  m\psi^2\frac{\partial}{\partial \psi^a}\right)u - 2 e^{mu}\psi_a
\equiv \hat\nabla^Q_a u- 2 e^{mu}\psi_a\;,\nn
&& \nabla^S_a
u=e^{-mu}\left[\frac{\partial}{\partial\psi^a} +
e^{4mu}\left(\psi^b\partial_{ab}+3m\psi^2D_a\right)\right]u
\;.\label{covder}
\eea
Here
\be
D_a = \frac{\partial}{\partial\theta^a}+\theta^b\partial_{ab} \;, \quad
\{D_a,
D_b\} = 2\partial_{ab}~,
\ee
is the standard covariant spinor derivative of $N=1, d=3$ Poincar\'e
supersymmetry. The Goldstone superfield $\lambda^{ab}$ is related to
$\Lambda^{ab}$ like in the bosonic case. Actually, in what follows we shall
need only the ``semi-covariant'' derivatives \p{covder}, and the superfield
$\lambda^{ab}$ will never appear in further consideration.

What we have at this stage, is a nonlinear realization of the $N=1$ AdS$_4$
supergroup  on the $N=2, d=3$ Goldstone superfield $u(x, \theta, \psi)$:
\be
\delta^* u(x,\theta,\psi) = -\left(\delta x^{ab}\partial_{ab} +\delta
\theta^a\partial^\theta_a +\delta \psi^a\partial_a^\psi\right)u(x,\theta,
\psi) + 2\theta^a\eta_a~, \label{transfact}
\ee
where $\delta^*$ means the ``active'' form of the infinitesimal
transformations. The first component in the $\theta, \psi$ expansion of
$u$ can be regarded as the Goldstone dilaton field discussed in the previous
Section. The spinor derivative $D_a u$ is shifted by $\eta_a$
under the $S$ -supersymmetry, suggesting that we actually face the $1/2$
spontaneous breaking of the AdS$_4$ supersymmetry, with
$D_a u|_{\psi=0}$ as the
corresponding Goldstone fermionic $N=1$ superfield. However, $u$ contains
quite a few extra component fields having no immediate Goldstone
interpretation.

To construct the minimal Goldstone multiplet, we shall generalize the method
which was applied in \cite{IKZL} to $d=2$ PBGS systems and, in \cite{Iv}, to
the   case of flat-space $N=1, D=4$ supermembrane. Following the reasonings
of
\cite{Iv} and keeping  in mind that the minimal scalar multiplets of $N=1$
AdS$_4$ supergroup are  represented by chiral $N=1, D=4$ or $N=2, d=3$
superfields, \footnote{From the structure relations \p{n1ads} it is seen
that
the complex combinations of spinor generators $Q_a + iS_a$ or $Q_a - iS_a$
form closed subgroups together with  the bosonic $SO(1,3)$ generators
$K_{ab},
M_{cd}$, thus showing the existence of two conjugated chiral subspaces in
the
$OSp(1|4)/SO(1,3)$ superspace $(x^{ab}, \theta^a, \psi^a, q)$ \cite{sorin}.}
we  shall regard the Goldstone superfield $u(z)$ to be {\it complex} and
subject it to the covariant  chirality constraint
\be\label{basiccon}
\left(\nabla_a^Q  - i\nabla_a^S\right) u = 0~.
\ee
This condition is equivalent to the similar one in terms of the full
covariant derivatives \p{fullcovder} in view of the relation
\be
\nabla_a^Q\pm i\nabla_a^S={{\delta_a^b \mp i\lambda_a^b}\over{
  \sqrt{1-2\lambda^2}}}\left( {\tilde\nabla}_b^Q\pm
           i{\tilde\nabla}_b^S \right)~,
\ee
and so it is covariant with respect to the whole $OSp(1|4)$.
Note that the transformation law \p{transfact} of the complex superfield $u$
cannot be rewritten in an equivalent passive form, because $\delta x^{ab},
\delta \theta^a, \delta \psi^a$ defined by eqs. \p{brokensusy} are
essentially complex due to the presence of $u(z)$.\footnote{Such a geometric
equivalent form of this transformation is expected
to exist in a complex chiral basis where \p{basiccon} become Grassmann
Cauchy-Riemann condition. Here we shall not elaborate on this point.}
Nevertheless, the transformations \p{transfact} are still self-consistent
just because of complexity of $u$ and can be checked to have the correct
closure.

Using the explicit form of the covariant derivatives \p{covder}, it is
straightforward to check that \p{basiccon} is self-consistent, in the sense
that the appropriate integrability condition is satisfied
\bea
&& \{\hat\nabla^Q_a -i\nabla^S_a, \hat\nabla^Q_b -i\nabla^S_b\}u -
2 \left(\hat\nabla^Q_a -i\nabla^S_a\right)e^{m u}\psi_b -
2 \left(\hat\nabla^Q_b -i\nabla^S_b\right)e^{m u}\psi_a  \nn
&& = T_{ab}^c\left( \nabla^Q_c -i\nabla^S_c\right)u~. \label{int}
\eea

The constraint \p{basiccon} can be solved by expanding $u$ in powers of
$\psi^a$. It is easy to check that \p{basiccon} expresses all terms in this
expansion in terms of $u_0 \equiv u|_{\psi^a = 0}$ and $D_a$ derivatives of
$u_0$. In particular,
\be
\frac{\partial u}{\partial\psi^a}|_{\psi=0}=- ie^{2mu}D_a u|_{\psi=0}~.
\ee
Thus the complex $N=1, d=3$ superfield
\be
u_0(x, \theta) \equiv q(x, \theta) + i \Phi(x, \theta)~, \quad q^\dagger =
-q~, \;\Phi^\dagger = - \Phi~, \label{defqPhi}
\ee
incorporates the full irreducible field content of the $N=2, d=3$ Goldstone
chiral superfield $u(x, \theta, \psi)$. Its $S$-supersymmetry transformation
reads \be
\delta u_0= L u_0+2\eta^a\theta_a + ie^{2mu}\eta^aD_a u_0 \;, \label{urule}
\ee
where
\be
L \equiv -m\left( \theta^2\eta^a+4x^{ab}\eta_b \right)
 \frac{\partial}{\partial\theta^a}+ 4m\eta_c\theta^bx^{ac}\partial_{ab}\;.
\ee
For the imaginary and real parts of $u_0$ eq. \p{urule} implies the
following
transformation rules
\bea\label{susy2}
&& \delta q= Lq -e^{2mq}\eta^a \left[ \sin(2m\Phi) D_a q+
   \cos(2m\Phi)D_a\Phi\right] +2\eta^a\theta_a \; , \nn
&& \delta \Phi= L\Phi +e^{2mq}\eta^a \left[ \cos(2m\Phi) D_a q-
   \sin(2m\Phi)D_a\Phi\right]\; .
\eea

The nonlinear realization we have at this step is still non-minimal in
the following sense. Besides the $N=1$ superfield $q(x,\theta)$
which contains all Goldstone fields required by the $1/2$ breaking of
$OSp(1|4)$ down to its $N=1, d=3$  Poincar\'e subgroup ($q|_{\theta = 0}$
for
the dilatations, $(D_a q)|_{\theta = 0}$ for the broken $S$-transformatons
and
$\partial_{ab}q|_{\theta = 0}$ for the broken $SO(1,3)/SO(1,2)$
transformations), there is an extra non-Goldstone $N=1, d=3$ superfield
$\Phi(x,\theta)$. The last step is to eliminate the latter in terms of $q$
and
its derivatives by imposing  some nonlinear covariant constraint on
$u_0(x,\theta)$,  analogous to the constraints imposed in the flat case
\cite{IK1}. The precise form of such a constraint can be found by applying
the
general procedure of refs. \cite{IKap,DIK,IKZL} to the given case. It goes
straightforwardly,  but it is rather technical in view of more
nonlinearities
involved as compared to the flat case. So we skip details and present the
final form of the constraint
\be\label{lastcon}
\Phi={ {e^{2mq}D^a q D_a q}\over{4+e^{2mq} D^2\Phi}}~.
\ee
It can be directly checked to be covariant with respect to
\p{susy2}.

{}From our superfield $u_0$ we can construct the following two simplest
invariants (up to normalization factors)
\be\label{S1}
S_1=\frac{1}{2}\int d^3x d^2\theta \left( e^{-4mu_0}+e^{4m u^\dagger_0}
\right) =    \int d^3x d^2\theta e^{-4mq}\cos(4m\Phi) \;,
\ee
and
\be\label{S2}
S_2= -\frac{1}{2im}\int d^3x d^2\theta \left( e^{-4mu_0}-e^{4m{u^\dagger_0}}
\right) =   {1\over m}\,\int d^3x d^2\theta e^{-4mq}\sin(4m\Phi) \;.
\ee
After solving the constraint \p{lastcon}
\be\label{Phisolv}
\Phi={ {e^{2mq}D^a q D_a q}\over{2+\sqrt{4+e^{4mq} D^2(D^b q D_b q)}}}~,
\ee
in the Lagrangians in $S_1$ and $S_2$ only lowest terms in $\Phi$ survive
in view of the nilpotency of $\Phi$ in \p{Phisolv}. So the actions take
the form
\be\label{S1a}
S_1 \sim \int d^3x d^2\theta e^{-4mq}\;,
\ee
\be\label{S2a}
S_2 \sim \int d^3x d^2\theta{ {e^{-2mq}D^a q D_a q}\over{2+
  \sqrt{4+e^{4mq}D^2(D^b q D_b q)}}}~.
\ee
Relevant for our purposes is just $S_2$, because it contains the kinetic
term
of the Goldstone superfield $q(\zeta)$. Its bosonic part, with the fermions
omitted and the auxiliary field $B = D^2q|_{\theta=0}$ eliminated by its
equation of motion ($B = 0$ in the vanishing fermions limit),
is recognized as the AdS$_4$ membrane action \p{baction2}.

We come to the conclusion that the Goldstone superfield action \p{S2a} is
the natural superextension  of the conformally-invariant AdS$_4$ membrane
action. Besides being manifestly invariant under $N=1, d=3$ Poincar\'e
supersymmetry, it is invariant under the nonlinearly realized  part of
$N=1$ AdS$_4$ supersymmetry $OSp(1|4)$ which acts on the $N=1, d=3$
superworldvolume as the Goldstone superfield-modified $d=3$ superconformal
transformations (eqs. \p{susy2} with \p{Phisolv} standing for $\Phi $).
Thus it is a PBGS superfield form of the worldvolume action of $N=1$
AdS$_4$ supermembrane. In the limit $m \rightarrow 0$, it reproduces the
known PBGS action of $N=1, D=4$ supermembrane in the flat Minkowski
background \cite{IK1}.

Finally, let us comment on the effect of adding extra invariant \p{S1a}.
The bosonic part of the  modified action $S_2+\alpha m^{-1}
S_1\,, \alpha^\dagger = \alpha\,,$ reads
\be
S_2^{bos}{\;}' \sim \int d^3x\left[
  e^{-6mq}\left( 1- \sqrt{1-{1\over 4}e^{4mq}\left( 2 \partial
q\cdot\partial
q+ B^2\right)}\,\right) +2\alpha e^{-4mq} B \right] \;. \ee
After elimination of $B$ by its algebraic equation of motion, the bosonic
action, again up to an overall renormalization factor, becomes
\be
S_{2}^{bos}{\;}'\sim \int d^3x e^{-6mq}\left({1\over \sqrt{1 + 16\alpha^2}}
-  \sqrt{1-{1\over 2}e^{4mq} \,\partial q\cdot\partial q}\, \right).
\ee
Thus, the effect of non-zero parameter $\alpha$ amounts to the appearance of
a sort of ``cosmological'' term on the worldvolume of the AdS$_4$ membrane
(the
structure of fermionic terms is also changed). The possibility of adding
such
an invariant term exists already in the flat case \cite{IK1,Iv1}, where
$\alpha m^{-1}S_1 \rightarrow -\alpha \int d^3xd^2\theta\, q(\zeta)$.

\section{Conclusions}
In this paper, proceeding from a $1/2$ partial breaking of the $N=1$ AdS$_4$
supersymmetry in the nonlinear realizations description, we have constructed
the worldvolume superfield action \p{S2a} for AdS$_4$ supermembrane. It is
the
first example of the complete PBGS Goldstone superfield action
for superbranes on curved superbackgrounds and, in particular, for AdS
superbranes. Its main characteristic feature is that the spontaneously
broken
part of $OSp(1|4)$ is realized as a Goldstone-superfield modified $d=3$
superconformal symmetry. Like in the case of flat supermembrane \cite{IK1},
the
superfield Lagrangian density of the AdS$_4$ supermembrane PBGS action is
not
of a tensor form, it is shifted by a full derivative under the broken
part of $OSp(1|4)$ transformations. In this sense, it resembles WZW or
Chern-Simons terms.

Our consideration here can be regarded as a first step towards constructing
analogous worldvolume superfield actions for more interesting examples of
branes on the superbackgrounds with the AdS$_n \times S^m$ bosonic part,
including the appropriate D$p$-branes. It still remains to be examined how
such  actions are related to the more familiar Green-Schwarz type ones.
Usually, the component on-shell form of PBGS actions coincides with a
static-gauge form of the appropriate G-S actions, with
$\kappa$-supersymmetry
also being properly fixed \cite{AGIT}. Their full off-shell superfield form
can be recovered from the superembedding approach \cite{emb1}. It would be
of
interest to establish similar relationships for PBGS actions of AdS branes,
in particular, for the action constructed here. Note that there is a problem
of the most convenient choice of the $\kappa $-symmetry gauge-fixing in the
worldvolume actions of AdS superbranes (see, e.g.,
\cite{padua,GSads2,kappa}).
The PBGS approach yields the superbrane actions at once in terms of the
physical worldvolume degrees of freedom (after elimination of  the auxiliary
fields, if they are present), therefore no problems related to the
non-uniqueness of the $\kappa$-gauge fixing can arise in this approach.

Finally, let us shortly comment on some related works.

The partial breaking of $D=4$ superconformal symmetries down to the
corresponding Poincar\'e supersymmetries in the nonlinear realizations
superspace framework was considered in \cite{NLRconf}. In these studies, the
superconformal symmetries are not regarded as AdS ones in higher dimensions
and their realization on the Poincar\'e superspace  coordinates has the
standard form  \cite{sconf} involving no nonlinear Goldstone superfield
terms
(appearing in our AdS realization). Respectively, the Goldstone superfield
actions of \cite{NLRconf} admit no superbrane interpretation.

In \cite{uematsu2}, in a similar nonlinear realizations setting, a partial
breaking of $N=2$ AdS supersymmetries in $D=3, 2$ to their $N=1$ AdS
subgroups
was considered. The basic Goldstone superfield was found to be associated
with an internal $U(1)$ symmetry generator. The corresponding invariant
Goldstone superfield action (yet to be constructed) also seems to bear no
direct links to AdS superbranes. Besides, it should be manifestly invariant
under $N=1$ AdS$_{3,2}$ supersymmetries and so explicitly include the
superspace coordinates, along the line of ref. \cite{sorin}. This is in
contrast with our action \p{S2a} which reveals manifest $N=1, d=3$
Poincar\'e
supersymmetry.

In a recent preprint \cite{kma}, a nonlinear realization of $N=1, D=4$
superconformal symmetry $SU(2,2|1)$ treated as $N=1$ AdS$_5$
supersymmetry was constructed. Conceptually, the approach of \cite{kma} is
close
to ours. However, the  invariant $N=1, d=4$ Goldstone superfield action
suggested there seems not to be the appropriate one to describe AdS$_5$
super 3-brane. Its Lagrangian (covariantized $N=1, d=4$ supervolume)
behaves as a density under the broken transformations, while the correct
minimal PBGS superbrane action is expected to be of non-tensor type, like
the
action constructed here.
\vspace{1.cm}

\noindent{\Large\bf Acknowledgements}\\

\noindent We are grateful to D. Sorokin, M. Tonin and B. Zupnik for
useful discussions. The work of E.I. and S.K. was supported in part
by the grants RFBR-CNRS 98-02-22034, RFBR-DFG-99-02-04022, RFBR
99-02-18417, INTAS-00-00254, NATO Grant PST.CLG 974874 and PICS Project
No. 593. E.I. thanks the Directorate of ENS-Lyon for the hospitality
extended to him during the course of this work.

\end{document}